\documentclass[aps,twocolumn,groupedaddress,showpacs,nofootinbib]{revtex4}

\usepackage{graphicx}% Include figure files
\usepackage{hyperref}
\usepackage{dcolumn}% Align table columns on decimal point
\usepackage{bm}% bold math
\usepackage{epsfig,amsbsy,bm,amssymb}
\usepackage{amsmath}  % needed for \tfrac, \bmatrix, etc.
\usepackage{amsfonts} % needed for bold Greek, Fraktur, and blackboard

\usepackage{xcolor}

\renewcommand{\k}{{k}}

\newcommand{\vs}{\vspace*}
\newcommand{\np}{\newpage}
\newcommand{\w}{\omega}
\newcommand{\W}{\Omega}

\newcommand{\eref}[1] {(\ref{#1})}
\newcommand{\Eref}[1] {Eq.~(\ref{#1})}
\newcommand{\Fref}[1] {Fig. \ref{#1}}

\newcommand{\Sref}[1] {Sec.~\ref{#1}}

\newcommand{\nn}{\nonumber}
\newcommand{\be}{\begin{equation}}
\newcommand{\ee}{\end{equation}}
\newcommand{\br}{\begin{eqnarray*}}
\newcommand{\er}{\end{eqnarray*}}
\newcommand{\ba}{\begin{eqnarray}}
\newcommand{\ea}{\end{eqnarray}}
\newcommand{\bp}{\begin{minipage}}
\newcommand{\ep}{\end{minipage}}
\newcommand{\bt}{\begin{tabular}}
\newcommand{\et}{\end{tabular}}

\newcommand{\ms}{\vspace*{-5mm}}
\newcommand{\mms}{\vspace*{-2.5mm}}
\newcommand{\st}{\small\tt}
\renewcommand{\l}{\lambda}

\renewcommand{\k}{{\bm k}}

\newcommand{\x}{{\bm x}}
\newcommand{\y}{{\bm y}}

  \newcommand{\R}{{\bm R}}

  \newcommand{\A}{{\bm A}}

\renewcommand{\t}{\tau}
\renewcommand{\l}{\lambda}

\renewcommand{\H}{H$_2$~}
\newcommand{\Hp}{H$_2^+$~}

\newcommand{\Wcm}[2]{
$\rm {#1}\times10^{{#2}}~W/cm^2$}

\begin{document}
\bibliographystyle{apsrev}
\title{XUV ionization of the \H molecule studied with 
attosecond angular streaking}

\author{Vladislav ~V. Serov$^{1}$}
\author{Anatoli~S. Kheifets$^{2}$}

\affiliation{$^{1}$General, Theoretical and Computer
  Physics, Saratov State University, Saratov
  410012, Russia}

\affiliation{$^{2}$Research School of Physics, The Australian National
  University, Canberra ACT 2601, Australia} 
\email{A.Kheifets@anu.edu.au}

 \date{\today}


\begin{abstract}
We study orientation and two-center interference effects in attosecond
time-resolved photoionization of the \H molecule. Time resolution of
XUV ionization of \H is gained through the phase retrieval
capability of attosecond angular streaking demonstrated earlier by
Kheifets {\em~et~al} [arXiv:2202.06147 (2022)].  Once applied to \H,
this technique delivers an anisotropic phase and time delay which both
depend sensitively on the molecular axis orientation. In addition, the
photoelectron momentum distribution displays a very clear two-center
interference pattern. When the interference formula due to Walter and
Briggs [J. Phys. B {\bf 32} 2487 (1999)] is applied, an effective
photoelectron momentum appears to be greater than the asymptotic
momentum at the detector. This effect is explained by a molecular
potential well surrounding the photoemission center.
 \end{abstract}

\maketitle
%\end{document} %stop

Attosecond time resolved studies of molecular photoionization have
become a rapidly growing field. Starting from the pioneering
experiment of \citet{PhysRevLett.117.093001} on H$_2$O and N$_2$O, the
method of attosecond interferometry has been progressively used
combining an extreme-ultraviolet (XUV) attosecond pulse train (APT)
and a synchronized infrared (IR) pulse. This technique has also been
known as reconstruction of attosecond beating by interference of
two-photon transitions (RABBITT) \cite{MullerAPB2002,TomaJPB2002}.
Recent applications of RABBITT to molecular photoionization include
attosecond resolution of coupled electron and nuclear dynamics in
dissociative ionization of \H \cite{Cattaneo2018} and
orientation-dependent time delay and electron localization studies in
CO \cite{Vos2018}. \citet{Nandi2020} resolved attosecond timing of
electron emission from a shape resonance in
N$_2$. \citet{PhysRevA.102.023118} recorded electron correlation
effects in attosecond photoionization of
CO$_2$. \citet{PhysRevA.104.063119} explored the role of
nuclear-electronic coupling in attosecond photoionization of \H.

The roadmap of atomic and molecular physics \cite{Young2018} has
identified X-ray free-electron lasers (XFELs) as a promising tool for
resolving ultrafast molecular dynamics. Attosecond time-energy
structure of XFEL pulses has been recently demonstrated
\cite{Hartmann2018,Duris2020}. This demonstration makes XFEL sources
potentially suitable for attosecond time resolution of atomic and
molecular photoionization. The only stumbling block preventing such an
application is a stochastic nature and an inherent time jitter of XFEL
radaition.

The method of attosecond angular streaking of XUV ionization was
developed to overcome this obstacle.  Prompted by theoretical works
\cite{Zhao2005,Kazansky2016,SiqiLi2018,Kazansky2019}, this method was
eventually implemented in practice for a shot-to-shot characterization
of isolated attosecond pulses (IAP) at XFEL
\cite{Hartmann2018,Duris2020}.  Angular streaking of XUV ionization
(ASXUVI or ASX for brevity) has common elements with the two
previously developed techniques: attosecond angular streaking known as
the attoclock \cite{Eckle2008,Eckle2008nphys,Pfeiffer2012} and the
attosecond streak camera (ASC)
\cite{Constant1997,ItataniPRL2002,Goulielmakis2004,Kienberger2004,Yakovlev2005,Fruhling2009,Zhang2011,PhysRevLett.107.213605}. 
As in ASC, ASX uses XUV pulses to ionize the target. Then, similarly
to the attoclock, the photoelectrons are steered by a circularly
polarized laser field which makes its imprint on the photoelectron
momentum distribution (PMD). This imprint is most visible in the plane
perpendicular to the laser propagation direction.  In its original
form
\cite{Kazansky2016,SiqiLi2018,Kazansky2019,Hartmann2018,Duris2020},
ASX employed an intense IR laser field and was interpreted within the
strong field approximation (SFA) \cite{XiZhao2022}. In these strong
field settings, the phase of the XUV ionization is usually neglected
and the timing information associated with this phase is lost.  An
alternative view within the lowest order perturbation theory (LOPT)
\cite{0953-4075-45-18-183001,Dahlstrom201353,Maquet2014} considers IR
streaking as an interference phenomenon which opens a natural access
to the streaking phase $\Phi_S$. The latter is typically decomposed
into the XUV ionization phase (or Wigner phase) and the
continuum-continuum (CC) phase from the IR interaction. These two
phases can be converted to the corresponding time delay components,
which add up to the atomic time delay $\tau_a$.

Phase retrieval capability of ASX based on this analysis was demonstrated
recently by \citet{Kheifets2022}. In their numerical simulations on
the hydrogen atom, they recovered accurately the streaking phase and
the atomic time delay across a wide range of photon energies starting
from the threshold and exceeding it many times. Most
importantly, this phase retrieval could be conducted from a single XUV
shot. This is a significant advantage over the existing
interferometric techniques which require a systematic and controllable
variation of the XUV/IR pulse delay in one set of measurements in
order to record a streaking spectrogram or a RABBITT trace. This
recording require a precise and stable temporal synchronization of the
XUV/IR pulses which is not feasible at XFEL at present.

In this paper, we extend ASX to molecular photoionization. We solve
numerically the time-dependent Schr\"odinger equation (TDSE)
describing the hydrogen molecule driven by a combination of the
linearly polarized XUV and circularly polarized IR pulses. In our
simulations, the XUV/IR pulse delay is incremented in several
steps. By augmenting the isochrone analysis proposed by
\citet{Kazansky2016} with the energy dependent XUV ionization phase,
we are able to interpret the molecular TDSE results in terms of the
atomic time delay. While the phase and time delay determination is
most accurate combining several increments of the XUV/IR delay, the
accuracy is not significantly compromised with just a single XUV/IR
pulse delay.
We make a comparison with the previous RABBITT simulations on \H
\cite{Serov2017} and confirm validity of our interpreatation and
accuracy of our numerical results. We also demonstrate a strong
dependence of the time delay on the molecular axis orientation
discovered earlier in \Hp ion
\cite{PhysRevA.90.013423,PhysRevA.93.063417}.

The paper is organized into the following sections. In \Sref{S1}
we outline basics of the ASX method. In \Sref{S2} we describe our
computational procedure. In \Sref{S3} we analyze and interpret our
numerical results. In \Sref{S4} we give our concluding remarks.

\section{Basic considerations}
\label{S1}

The proposed phase retrieval by ASX is outlined in our preceding work
\cite{Kheifets2022}.  The basics of the molecular ASX are essentially
the same as for atoms. We proceed as follows. We apply the SFA and
write the photoionization amplitude as \cite{Kitzler2002}
\be
    a(\k ,\tau) = i \int_{t_0}^{\infty} \!\! dt \ E_x (t-\tau) 
 D_x\left[\k -\A (t)\right]  e^{-i\Phi(t)}
\ .
\ee
Here the electric field of the XUV pulse $E_x$ is advancing the
streaking pulse by the time $\t$. The streaking field is described by
its vector potential 
$$
\A (t) = 
A_0\cos(\omega t)\hat{\x} +A_0\sin(\omega t)\hat{\y}
\ .
$$ 
The photoelectron momentum is confined
to the polarization plane  $\k = k\cos\phi~\hat{\x} +
k\sin\phi~\hat{\y}$, where $\phi$ is the emission angle. 

The exponential term contains the  phase factor
\be
\Phi(t)=
\frac12\int_{t}^{\infty} dt^{\prime} 
\left[\k  - \A (t^{\prime})\right]^2 - E_0 \, t
\ ,
\ee
which contains  the photoelectron energy in the absence of streaking
 $E_0=\W-I_p$. 
The most probable photoelectron trajectory, starting
at the time $t_{\rm st}$,  keeps the phase stationary:
\be
\label{SFA} 
\Phi'(t_{\rm st}) = \frac12 |\k-\A(t_{\rm st})|^2-E_0=0 
\ee 
We assume that the XUV pulse is short relative to the IR pulse and
shifted relative to its peak position by the time $\t$. Under these
conditions, \Eref{SFA} is transformed to the following {\em~isochrone}
equation \cite{Kazansky2016}:
\be
\label{iso}
k^2/2-E_0 = kA_0\cos(\phi-\w\t) 
\ee
Here  we neglect the ponderomotive energy
$U_p=A_0^2/2$ in a weak streaking field.

The above stationary phase analysis should be modified to account for the
photoelectron energy dependence of the dipole matrix element
\cite{M.Schultze06252010}
\be
\arg  \left\{D\left[\k -\A (t)\right]\right\} 
\propto \alpha |\k-\A (t)|^2/2
\ ,
\ee
where
\vs{-0.5cm}
\be
\label{alpha}
\alpha = \partial \arg D(\sqrt{2E})/ \partial E
\ee
The modified stationary phase equation reads
\be 
\frac12 \left|\k - \A (t_{st})\right|^2 - E_0 +
\frac{\alpha}{2}\frac{d}{dt}\left[ \left(\k - \A (t_{\rm st})\right)^2 \right]
= 0 
\ee
This  leads to a generalized isochrone equation
\begin{eqnarray}
\label{isom}
k^2/2 - E_0 &=& kA_0 
\left[ \cos(\phi - \w \t) 
- \alpha\w\sin(\phi - \w \t) \right]
% \nn \\ &=& kA_0 \sqrt{1+\w^2\alpha^2}  
%\cos[\phi - \w t_{\rm st} +  \tan^{-1} (\w\alpha) ]
 \nn \\ &\approx & kA_0
\cos[\phi - \w \t +  \w\alpha ]
\end{eqnarray}
Here $\alpha=\Phi_S/\w=\t_a$ under certain XUV and IR pulse parameters
as demonstrated in \cite{Kheifets2022}.

\section{Computational details}
\label{S2}

%VVS
We solve numerically the molecular TDSE equation using the computer
code \cite{PhysRevA.84.062701} to obtain the ionization amplitude
$f(\k)$. We use an angular basis that included spherical harmonics up
to $l_{max}=7$ and $|m_{max}|=7$.  Unlike the dipole selection rules
in atomic XUV photoionization, the quantum numbers $l,m$ adhere to the
parity conservation.

The photoelectron momentum spectrum $P(\k)$ is obtained as the modulus
squared of the ionization amplitude
\be
P(\k) \propto |f(\k)|^2 
\ .
\label{f_expan_spheroid}
\ee
The PMD is restricted to the polarization plane $P(k_x,k_y,k_z=0)$
and converted to the polar coordinates $P(k,\phi)$ where
\be
\label{polar}
k=(k_x^2+k_y^2)^{1/2}
 \ , \
\phi = \tan^{-1}(k_y/k_x)
\ .
\ee
In these coordinates, we define the directional probability of the
photoelectron emission
\be
\label{directional}
P(\phi) = \int dk \  P(k,\phi)
\ee
and the mean (central) radial momentum in the given direction 
\be
\label{central}
\bar k(\phi) = \int k P(k,\phi) dk / P(\phi)
\ .
\ee

The TDSE is driven by the XUV and IR pulses with the following
parameters. The XUV pulse with a Gaussian envelope has a FWHM of 2~fs
and the intensity of \Wcm{6}{13}. The XUV photon energy $\W$ ranges
from 0.7~au to 3~au.  A relatively low XUV field intensity is required
to remain within the LOPT framework. A fairly large pulse duration is
employed to ensure a moderately narrow spectral width to probe XUV
ionization sufficiently close to the threshold at 15.6~eV
(0.57~au). At the same time, the spectral width $\Gamma$ should be
kept sufficiently large to make sure the IR assisted XUV absorption
process overlaps spectrally with unassisted XUV ionization
\cite{Kheifets2022}. This requires $\Gamma> 2\w$, where $\w$ is the
laser photon energy. To satisfy this requirement, we chose a mid-IR
laser pulse with $\w=0.038$~au corresponding to $\l=1200$~nm. The
pulse has a cosine squred envelope with  FWHM of
25~fs and the intensity of \Wcm{1.5}{11}. The XUV pulse is linearly
polarized along the $\hat\x$ axis whereas the IR pulse is circularly
polarized in the $(xy)$ plane.  At each XUV photon energy, we scan the
delay between the XUV pulse and the IR laser field ($\tau$) in the
range of 0 to 60 au in 7 increments.

\section{Numerical results}
\label{S3}

We identify three regions in the photoelectron energies which display
distinctively different PMD in the polarization plane. These regions
can be characterized by the strength of the molecular two-center
interference. The theory of this interference was proposed by
\citet{Cohen1966} and \citet{Kaplan69} and further developed for
diatomic molecules fixed in space by \citet{Walter1999}. In the latter
formulation, the ionization amplitude is approximated by the
expression
\be
\label{Fano}
f_{\rm WB}(\k)\propto ({\bm e} \cdot \k)\cos(\k\cdot\R/2)
\ ,
\ee
where $\bm e$ is the polarization vector of light and $\R$ is the
vector connecting the nuclei. The first term in the RHS of \Eref{Fano}
is the atomic hydrogen  dipole factor whereas the second term
represents the molecular two-center interference.  In the following,
we will use a scalar coefficient $c=kR/2$ to identify the strength of
this interference.

\begin{figure}[t]
\mms
\epsfxsize=8.5cm
%\epsffile{Vlad/0.7/x/3H.eps}
\epsffile{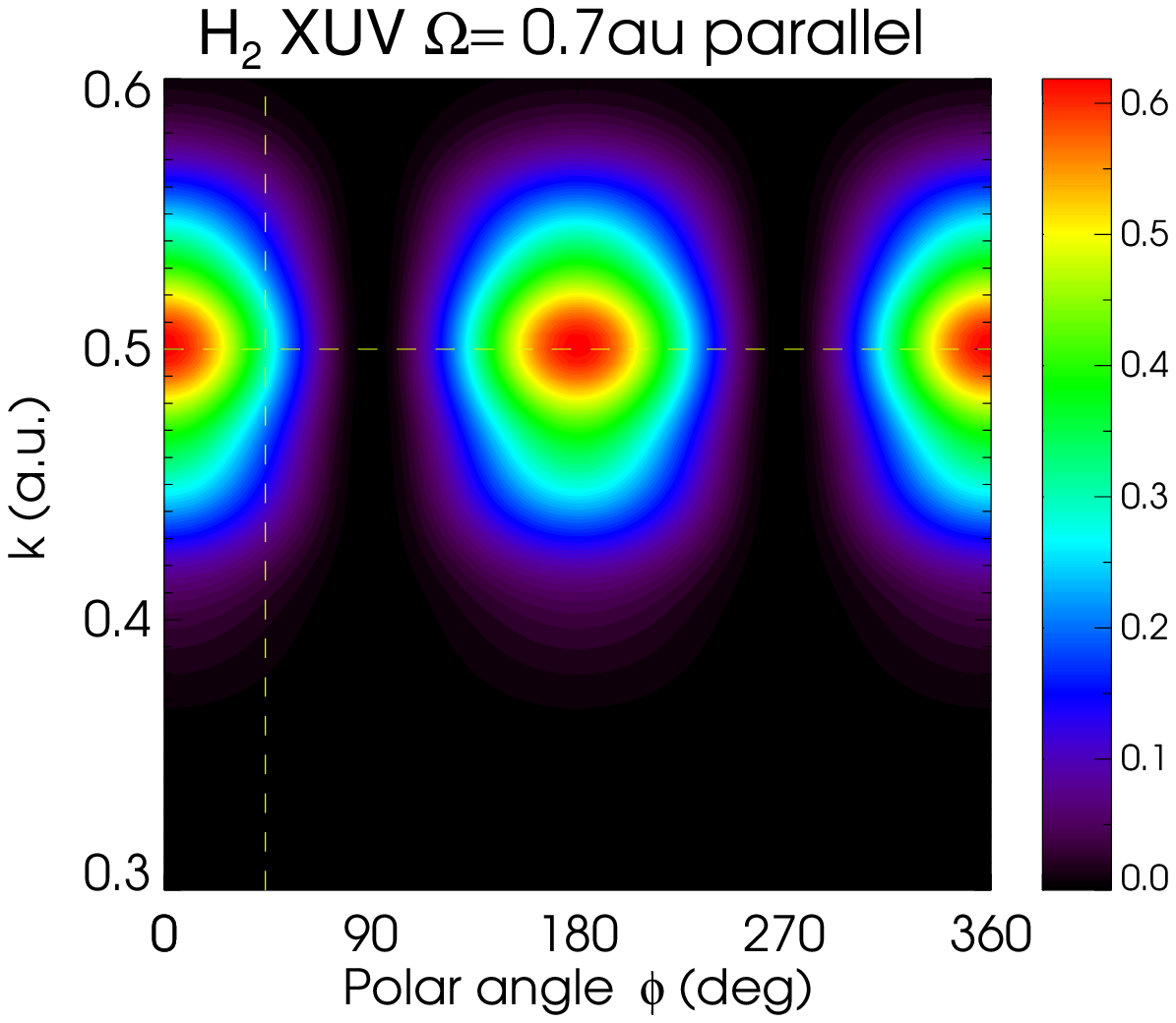}

\ms\ms
\epsfxsize=8.5cm
%\epsffile{Vlad/0.7/0/3H.eps}
\epsffile{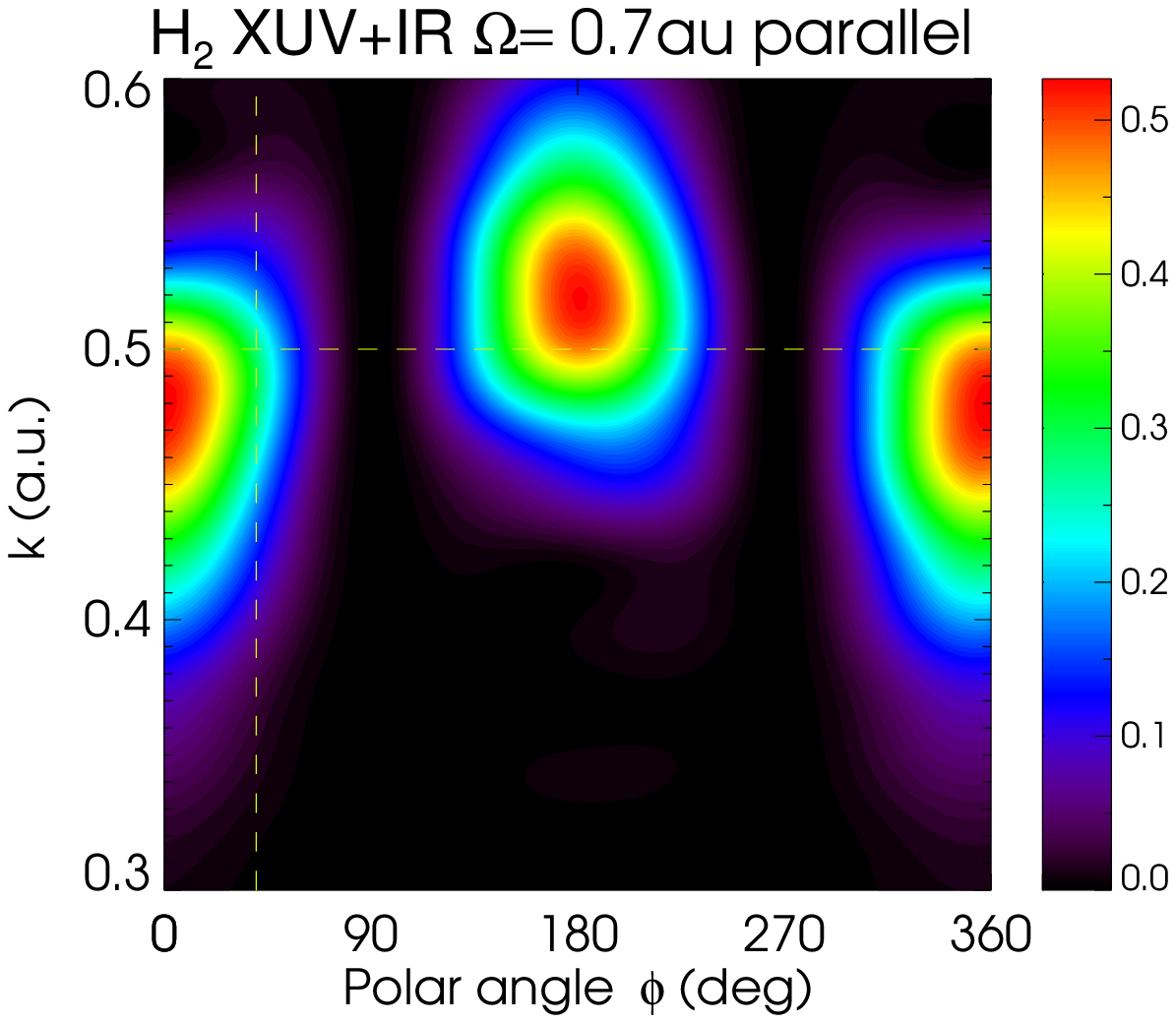}

\ms\ms
\epsfxsize=6cm
%\epsffile{Vlad/0.7/0/PLOTk.eps}
\epsffile{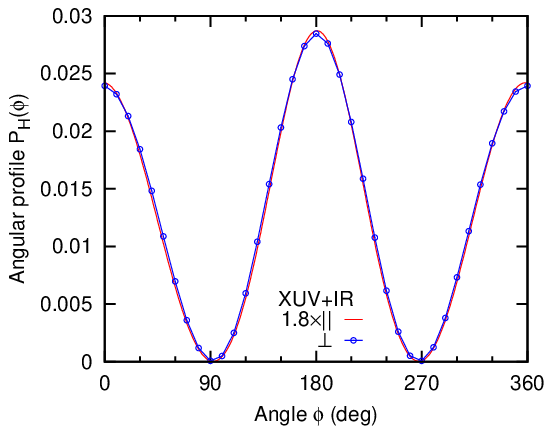}

\mms
\caption{Top:  PMD of \H at $\W=0.7$~au in the parallel field
  orientation with the XUV only pulse (top) and the XUV+IR pulses
  (middle). The horizontal dashed line visualize the photoelectron
  momentum  $k_0=\sqrt{2(\W-I_p)}$ from the energy conservation. The
  vertical dashed line mark the half  of the angular width.  
\label{Fig1}}
\ms
\end{figure}

\subsection{Weak interference}
\label{weak}

At low photoelectron energy when $c\ll1$, the PMD of \H looks
essentially atomic like with very little anisotropy seen between the
parallel and perpendicular orientation of the molecular axis $\R$
relative to the linear polarization axis $\bm e$ of the XUV pulse.
This behavior is featured in \Fref{Fig1} which displays the PMD at
$\W=0.7$~au. The top and middle panels both illustrate the case of the
parallel orientation with the XUV only pulse (top) and XUV+IR pulses
(middle). The bottom panel displays the radially integrated PMD of the
middle panel in the form of the angular distribution $P(\phi)$ which
is overlapped with the analogous distribution for the perpendicular
orientation. Except for an overall magnitude factor $\times1.8$, the
$\perp$ angular distributions look essentially the same as the
$\parallel$ one.

\begin{figure}[h]
\epsfxsize=6.5cm
%\epsffile{Vlad/0.7/PLOTh.eps}
\epsffile{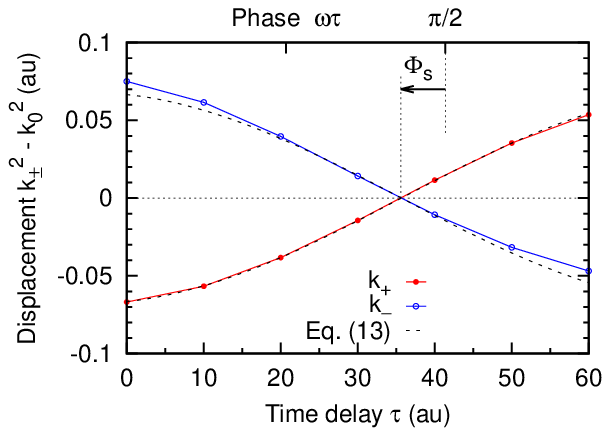}

\epsfxsize=6.5cm
%\epsffile{Vlad/0.7/PLOTH.eps}
\epsffile{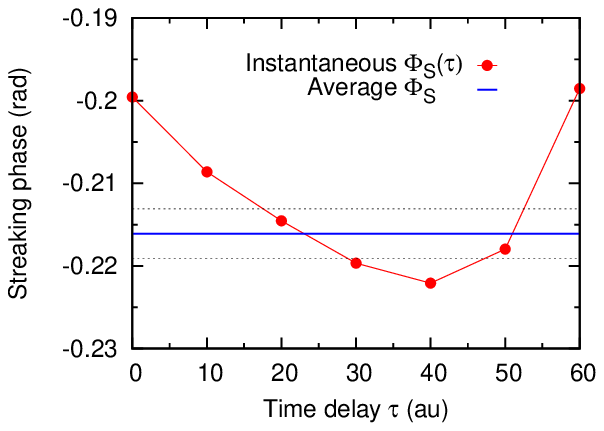}

\caption{Top:  Radial
  momentum displacements $k_\pm^2/2-k_0^2/2$ are shown at various
  XUV/IR delays $\t$. The dashed line represents the fit with
  \Eref{ansatz}. The arrow indicates the streaking phase $\Phi_S$.
Bottom: the fit with \Eref{ansatz} is applied  to individual
$\t$ values to determine the instantaneous $\Phi_S(\t)$. The average
$\Phi_s$ is shown as a solid line with error bars visualized
by  dotted lines. 
\label{Fig2}}
\end{figure}

Meanwhile, the PMD of the top panel (XUV only) and the middle panel
(XUV+IR) differ by a noticeable displacement of the radial momentum by
the vector-potential $A_{\st IR}$ of the streaking field. To quantify
this displacement, we use the central photoelectron momenta
\eref{central} in the  downwards (-) and upwards (+) shifted lobes
of the PMD
$$
k_- \equiv \bar k(\phi=0)
\ \ , \ \ 
k_+ \equiv \bar k(\phi=\pi)
\ ,
$$
These momenta $k_\pm(\t)$, which depend sensitively on the XUV/IR time
delay $\t$, are then used to obtain the isochrone phase offset:
\be
\label{ansatz}
k_\pm^2(\t)/2-E_0 = \pm A_0\, k_\pm(\t)\cos(\w\t+\Phi_S)
\ .
\ee
This determination is illustrated in the top panel of
\Fref{Fig2}. Here we determine $\Phi_S=-0.216\pm0.003$~rad by fitting
either of the $k_\pm(\t)$ branches with a common streaking phase value
over the whole set of the time delays $\t$. Alternatively, we can apply
\Eref{ansatz} to individual $\t$ values and to determine the
instantaneous $\Phi_S(\t)$. These values are displayed along with the
average streaking phase on the bottom panel of \Fref{Fig2}. Even
though the variation of $\Phi_S(\t)$  exceeds the error bars of
the average value, the accuracy of the instantaneous streaking phase
determination is not significantly compromised.

%%%%%%%%%%%%%%%%%%%%%%%%%%%%%%%%%%%%%%%%%%%%%%%%%%%%%%

%%%%%%%%%%%%%%%%%%%%%%%%%%%%%%%%%%%%%%%%%%%%%

\begin{figure}[t]
\mms
\epsfxsize=8.5cm
%\epsffile{Vlad/1.5/x/3H.eps}
\epsffile{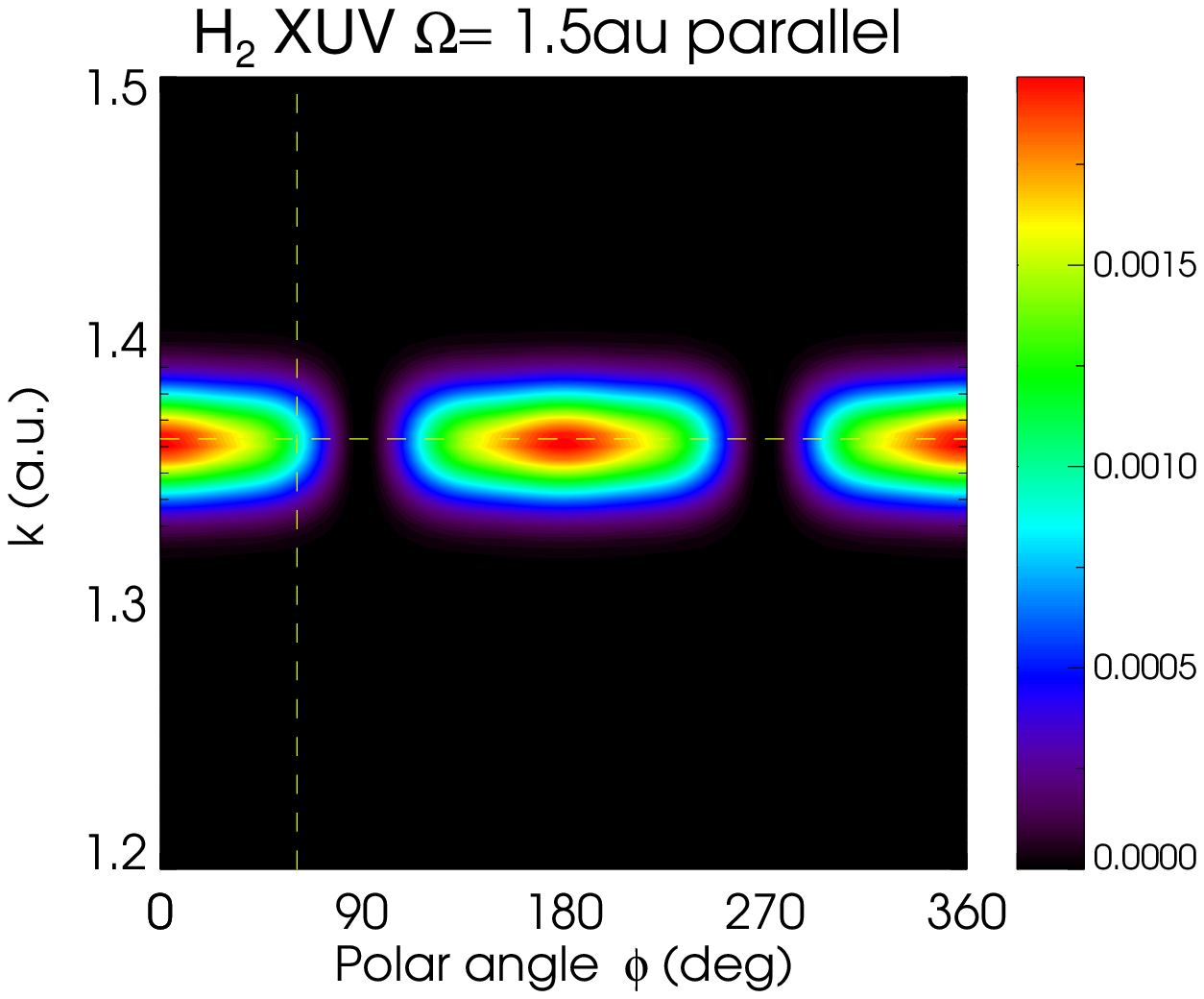}

\ms\ms
\epsfxsize=8.5cm
%\epsffile{Vlad/1.5/x/3Hp.eps}
\epsffile{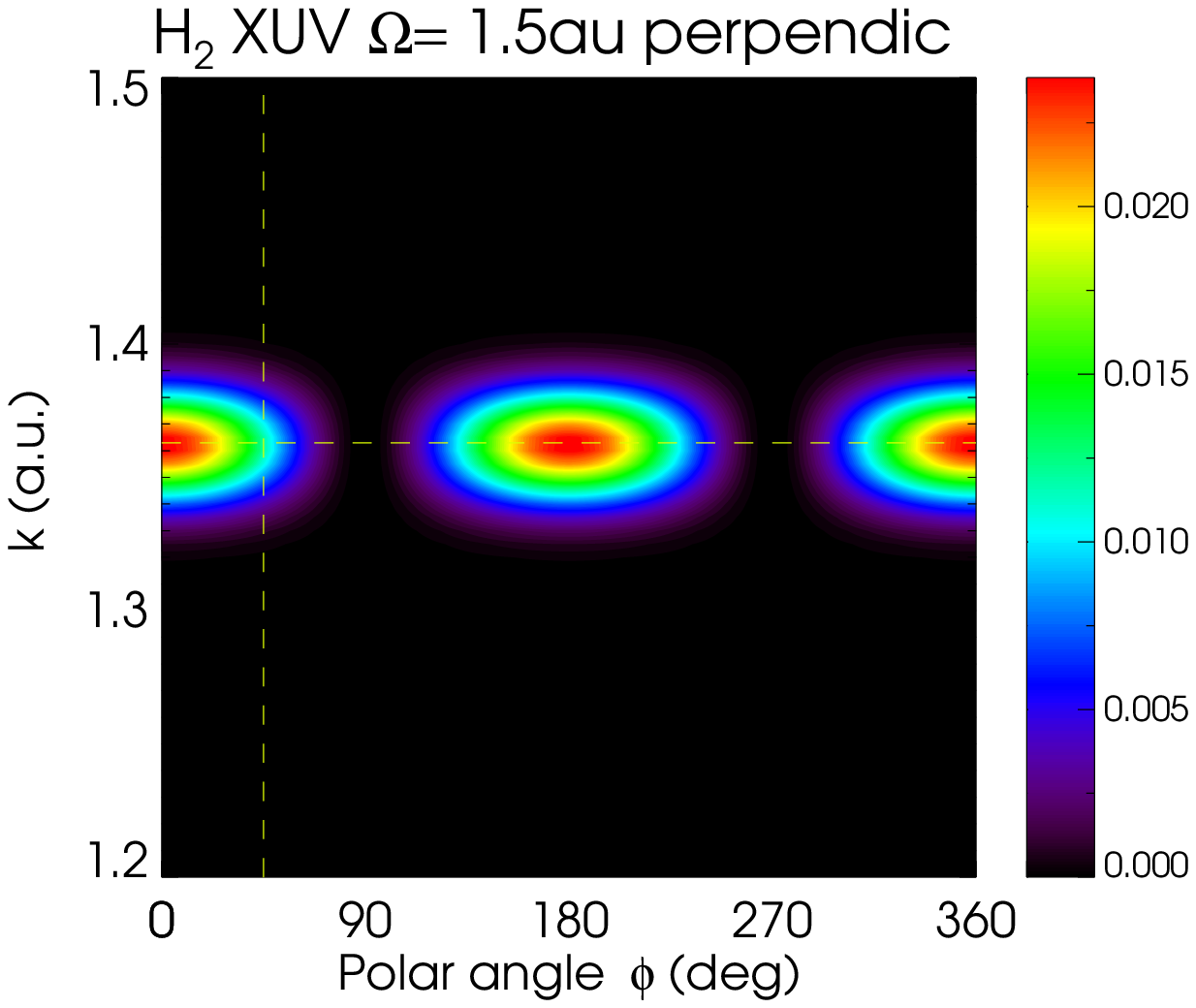}

\ms\ms
\epsfxsize=6cm
%\epsffile{Vlad/1.5/x/PLOTk.eps}
\epsffile{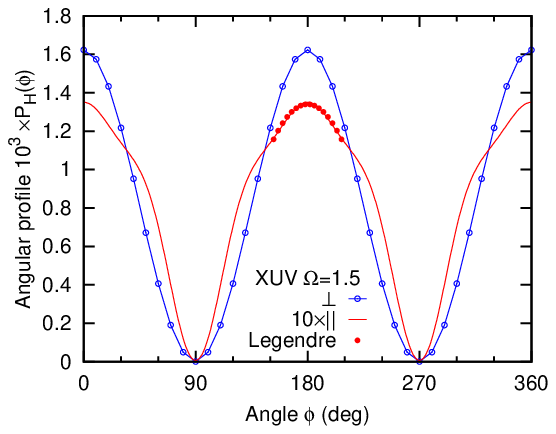}

\mms
\caption{Top: PMD of \H at $\W=1.5$~au for the parallel (top) and
  perpendicular (middle) field orientation with the XUV only pulse The
  horizontal dashed line visualize the photoelectron momentum
  $k_0$ while the vertical line marks half of the angular width.
\label{Fig3} }
\ms
\end{figure}

%%%%%%%%%%%%%%%%%%%%%%%%%%%%%%%%%%%%%%%%%%%%%%%%%%%%%%%

\subsection{Moderate interference}
\label{moderate}

This region is characterized by a moderate factor $c\lesssim 1$.  A
typical PMD in this region is presented in the top and middle panels
of \Fref{Fig3}. Here the XUV photon energy $\W=1.5$~au and the
molecule is oriented parallel (top) and perpendicular (middle) to the
polarization axis. Both panels visualize single-photon XUV
ionization. Adding the IR streaking field does not change the PMD
structure except for a vertical  up and down displacement by the
amount of $A_{\st IR}$ as in the middle panel of \Fref{Fig1}.

The case of $c\lesssim 1$ differs from $c\ll 1$ by a significant
deviation of the PMD shapes corresponding to the parallel and
perpendicular orientations. The PMD lobes are noticeably elongated for
the parallel orientation and acquire a greater angular width. The
photoelectron angular distribution shown in the bottom panel is
markedly different for the $\parallel$ and $\perp$ orientations. While
the latter retains the atomic like structure, the former widens
significantly and becomes drastically, by a factor $\times10$,
suppressed. This parallel emission suppression is documented in the
literature and termed the "confinement effect"
\cite{PhysRevLett.98.043005,PhysRevA.79.043409}. This corresponds to
the dominant photoelectron $p$-wave trapped inside a one-dimensional
box of length $R$ when the momentum quantization condition $kR=\pi$ 
satisfied at $c=\pi/2$.

\subsection{Strong interference}
\label{strong}

\begin{figure}[t]
\mms
\epsfxsize=8.5cm
%\epsffile{Vlad/2.5/x/3H.eps}
\epsffile{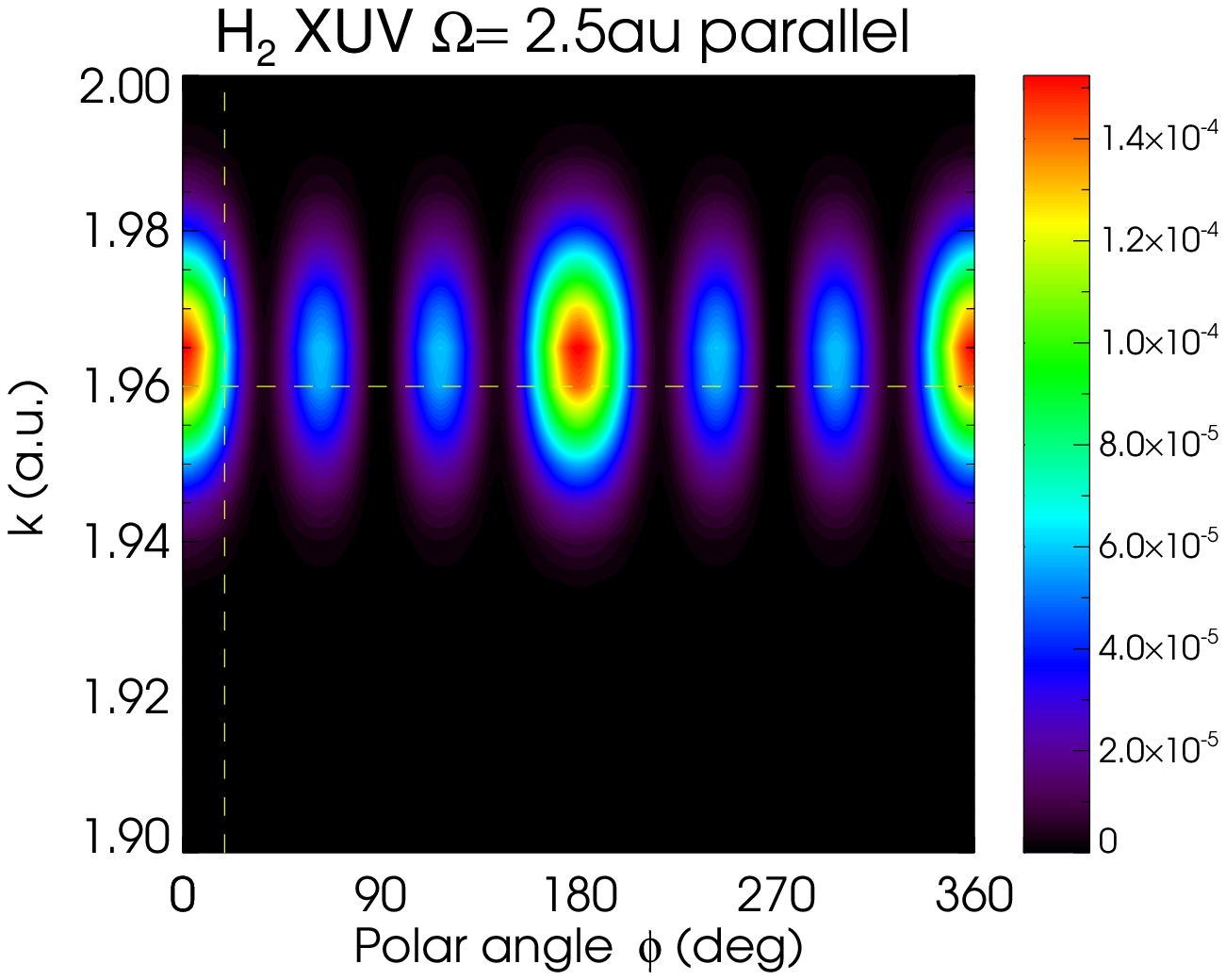}

\ms\ms
\epsfxsize=8.5cm
%\epsffile{Vlad/2.5/x/3Hp.eps}
\epsffile{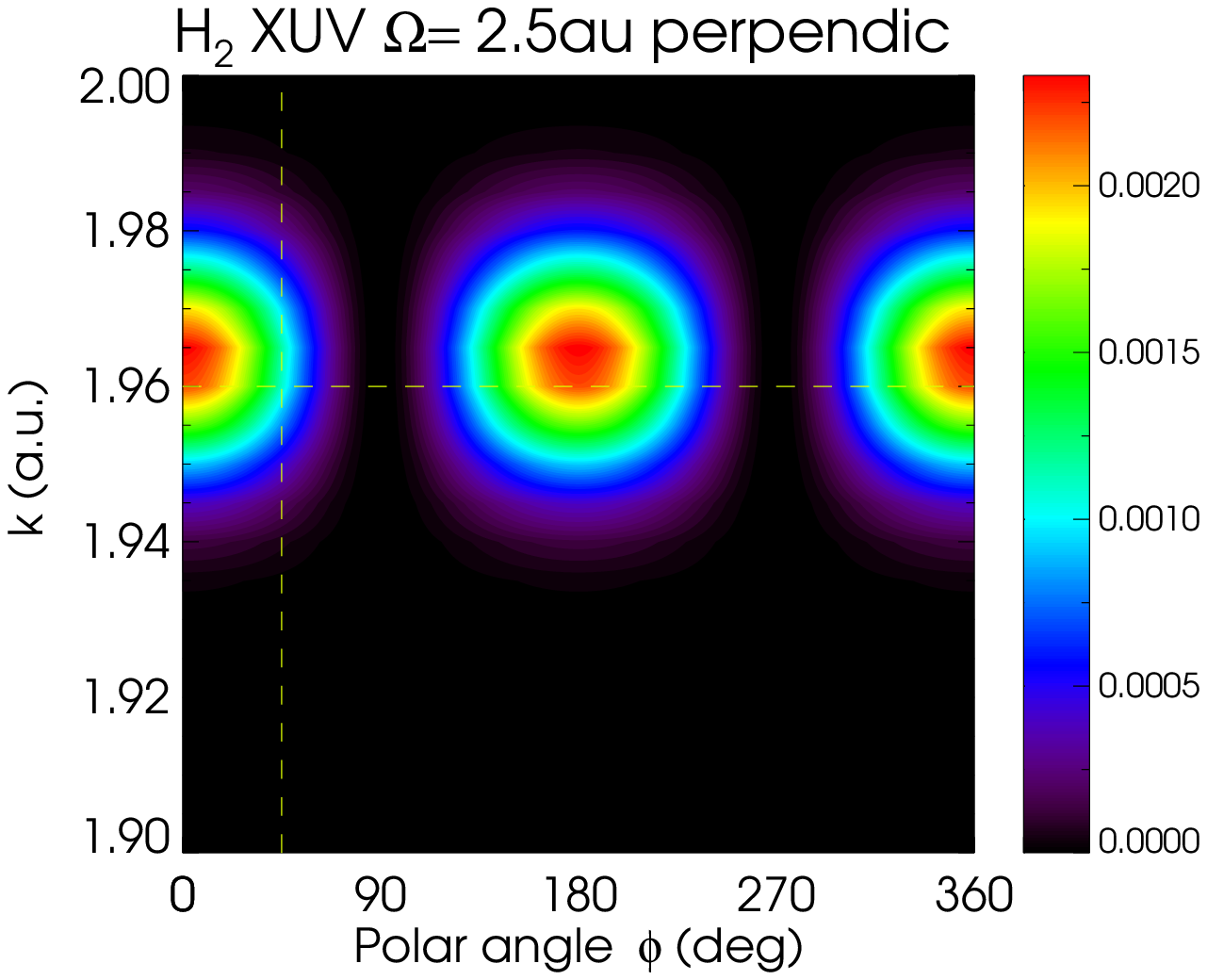}

\ms\ms
\epsfxsize=6cm
%\epsffile{Vlad/2.5/x/PLOTk.eps}
\epsffile{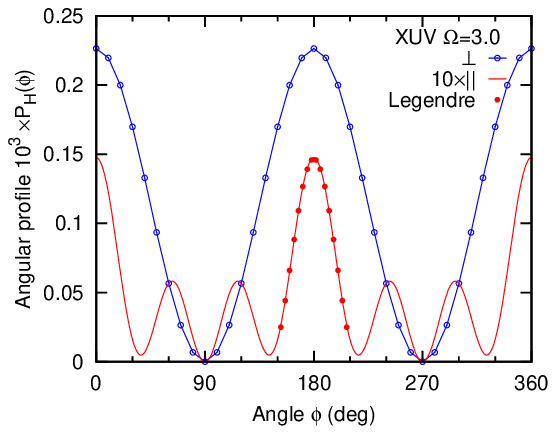}

\mms
\caption{Same as \Fref{Fig3} for $\W=2.5$~au  
\label{Fig4}}
\ms
\end{figure}

This region is characterized by a large interference factor
$c\gtrapprox \pi/2$.  In this region, the shape distortion of PMD is
most graphical as shown in \Fref{Fig4} for $\W=2.5$~au. While the
perpendicular orientation (middle panel) retains an atomic like shape,
the parallel orientation (top panel) displays very clear interference
fringes. These fringes are also seen in the angular resolved
cross-section exhibited in the bottom panel of  \Fref{Fig4}.

%%%%%%%%%%%%%%%%%%%%%%%%%%%%%%%%%%%%%%%%%%%%%%%%%%%%%%%
\begin{figure}[t]

\epsfxsize=7cm
%\epsffile{Vlad/PLOT5.eps}
\epsffile{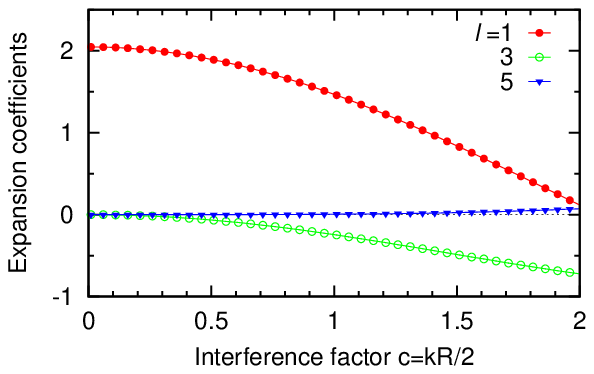}

\epsfxsize=7cm
%\epsffile{Vlad/PLOTw.eps}
\epsffile{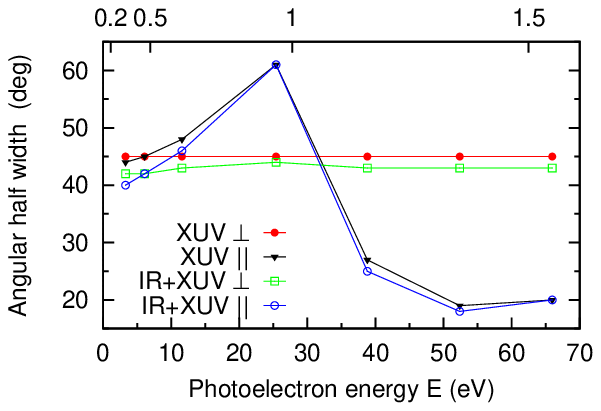}

\caption{Top: expansion coefficients of the ionization amplitude over
  the spherical harmonics \eref{Fano_exp} plotted as functions of the
  interference factor $c=kR/2$. Bottom: angular half width of the PMD
  lobes as a function of the photoelectron energy $E=\W-I_p$. The
  upper horizontal scale marks the corresponding interference factors.
\label{Fig5}}
\ms
\end{figure}

To quantify the two-center interference effects across a wide range of
the photon energies, we plot in the bottom panel of \Fref{Fig5} the
half width of the PMD lobes. The atomic like half width of $45^\circ$
corresponds to the dipole $\cos^2\phi$ angular shape. It is retained
consistently over the whole photon energy range in the perpendicular
molecular orientation for XUV only photoionization. Adding a streaking
IR field reduces this width insignificantly for the $\perp$
orientation. Meanwhile, the $\parallel$ orientation, both in XUV and
XUV+IR fields, displays a wide oscillation of the width in the range
of moderate to strong two-center interference. 

To understand the nature of this oscillation, we note that the
amplitude \eref{Fano} for the parallel orientation is reduced to
\be
\label{Fano1}
f^\parallel_{\rm WB}(\phi) \propto
\cos\phi \cdot \cos(0.5kR\cos\phi)
\ .
\ee
This amplitude can be expanded over the spherical harmonics with the
expansion coefficients given by the following expression
\cite{PhysRevA.86.025401}
\be
\label{Fano_exp}
A_{\ell}(c)= \left\langle Y_{\ell0} | f^\parallel_\text{WB}
\right\rangle = \sqrt{2\pi} \int_{-1}^{1} \bar{P}_{\ell}(\eta) \eta
\cos(c\eta) d\eta .  \ee
Here $\bar{P}_{\ell}(\eta)$ are the normalized Legendre polynomials
which depend on $\eta=\cos\phi$.  The expansion coefficients
\eref{Fano_exp} for various $\ell$ are plotted in the top panel of
\Fref{Fig5}.  From this graph we see clearly that $c\simeq1$
corresponds to a noticeable contribution of the $f$-wave whereas at
$c\simeq\pi/2$ the $p$- and $f$-wave contributions become of the same
magnitude. These two boundaries correspond to the region of moderate
and strong two-center interference according to our classification in
\Sref{moderate} and \Sref{strong}. In the meantime, the weak
interference $c\ll1$ considered in \Sref{weak} corresponds to a nearly
sole contribution of the $p$-wave.

Fitting the numerical TDSE results for the photoelectron angular
distributions with the squared amplitude \eref{Fano1} gives
systematically higher effective momenta $k_{\rm eff}$ in comparison
with the nominal momenta $k$ determined by the energy
conservation. We find $k_{\rm eff}$ from the moduli ratio of the
$f$- and $p$-waves
\be
\label{c_sol}
\left|\frac{A_{3}(c_a)}{A_{1}(c_a)}\right|= \left|\frac{\left\langle Y_{30} | f(\k) \right\rangle}{\left\langle Y_{10} | f(\k) \right\rangle}\right|
.
\ee
This ratio equates the expansion coefficients $A_{\ell}$ from
\Eref{Fano_exp} evaluated at  $c_a=k_{\rm eff}R/2$ with the
corresponding expansion coefficients of the exact numerical amplitude
$f(\k)$ found by the TDSE solution.

%VVS
The deviation $k_{\rm eff}$ from $k$ displayed in the top panel of
\Fref{Fig6} can be explained by the effective potential of the ion
remainder.  Due to this potential, the momentum of the electron near
the nucleus is greater, and, accordingly, a larger phase difference
between the emitting centers is accumulated. We can introduce an
average effective potential related to the effective momentum through
the following expression
\be
k_{\rm
  eff}/k=\sqrt{1+2|\bar{U}_{\rm eff}|/k^2}
\ .
\ee
The values of $\bar{U}_{\rm eff}$ are presented in the bottom panel of
\Fref{Fig6}.  A gradual reduction of $\bar{U}_{\rm eff}$ with a
decreasing XUV photon energy can be understood as follows. By the
uncertainty principle, a slower photoelectron has a larger birthplace
area across which the ionic potential is sampled. Therefore, its
effective depth becomes smaller.

\subsection{Streaking phase and time delay}

\begin{figure}
 \epsfxsize=6.5cm
% \epsffile{Vlad/PLOTk.eps}
 \epsffile{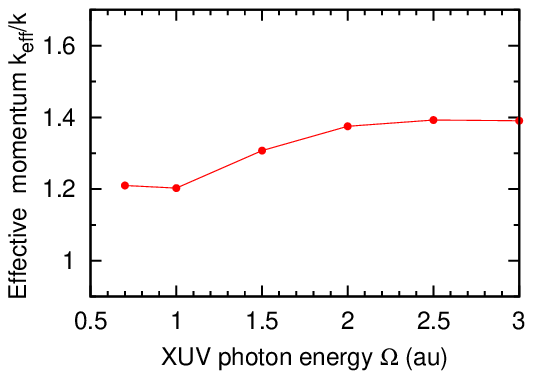}

 \epsfxsize=6.5cm
% \epsffile{Vlad/PLOTu.eps}
 \epsffile{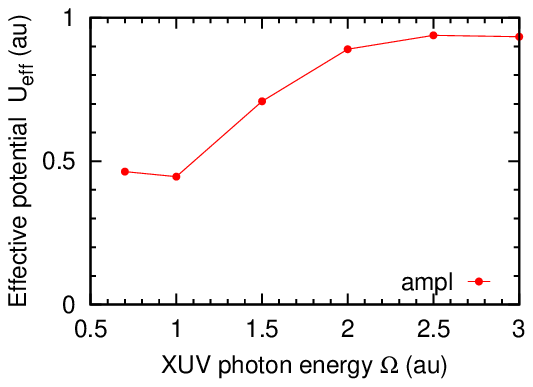}

  \caption{Top: Effective momentum $k_{\rm eff}/k$. Bottom: effective
    potential $U_{\rm eff}$.
\label{Fig6}}
\ms
\end{figure}

\begin{figure}
 \epsfxsize=8cm
% \epsffile{Vlad/PLOTp.eps}
 \epsffile{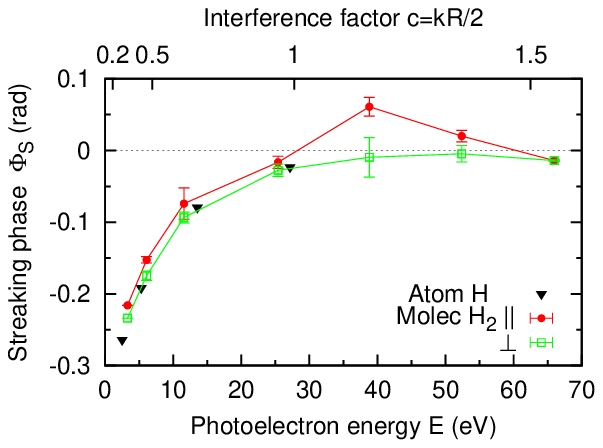}

 \epsfxsize=8cm
% \epsffile{Vlad/PLOTt.eps}
 \epsffile{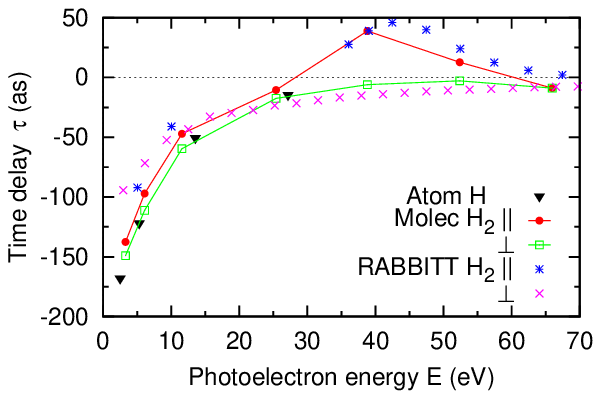}

  \caption{Top: Streaking phase $\Phi_S$ as a function of the
    photoelectron energy for the hydrogen atom and the \H molecule in
    the $\parallel$ and $\perp$ orientations. Bottom: the atomic time
    delay derived from the streaking phase $\t_a= \Phi_S/\w$ is
    compared with the corresponding values returned from the RABBITT
    simulations \cite{Serov2017}.
\label{Fig7}}
\ms
\end{figure}

The streaking phase results for the \H molecule in the $\parallel$ and
$\perp$ orientations are summarized in the top panel of \Fref{Fig7}
where they are compared with the corresponding values of the H
atom. While the molecular $\Phi_S$ in the $\perp$ orientation is very
similar to the atomic one, the $\parallel$ orientation displays a
systematically higher values, especially at the onset of the strong
interference when the $c$ factor approaching $\pi/2$. The atomic time
delay derived from the streaking phase $\t_a= \Phi_S/\w$ is shown in
the bottom panel of \Fref{Fig7} where it is compared with the
corresponding values returned by the RABBITT simulations
\cite{Serov2017}. Numerical $\t_a$ values from the ASX and RABBITT
simulations are slightly different because of a difference in the
wavelength $\l=1200$~nm in the former and 800~nm in the latter. The IR
photon wavelength and energy affect the CC component of the atomic
time delay \cite{Dahlstrom201353,Serov2015} which becomes particularly
noticeable close to the threshold. Nevertheless, the qualitative
behavior of $\t_a$ is very similar in both sets of simulations. The
atomic time delay in the H atom and the \H molecule in the $\perp$
orientation remain negative in the studied XUV photon energy range. At
the same time, the $\parallel$ orientation displays a sharp rise of
the time delay to positive values. This effect is also recorded in the
\Hp ion \cite{PhysRevA.90.013423,PhysRevA.93.063417}. It was
attributed in \cite{PhysRevA.90.013423} to the destructive two-center
interference. We offer a more physically appealing interpretation of
the positive time delay due to the trapping the photoelectron in the
molecular potential well. From the condition of this trapping $k_{\rm
  eff}R=\pi$ occurring at $kR\simeq2.4$ we can estimate $|U_{\rm
  eff}|\simeq 1$~au.  This determination is consistent with the values
of $U_{\rm eff}$ presented in the bottom panel of \Fref{Fig6}.

%
%% $$
%% \left({k\over k_0}\right)^2 = 1+ 2 {|U_{\rm eff}|\over k_0^2}
%% \ \ , \ \ 
%% 2 {|U_{\rm eff}|\over k_0^2} = \left({k\over k_0}\right)^2 - 1
%% \ \ , \ \ 
%% |U_{\rm eff}| = {k^2\over 2}-{k_0^2\over 2}
%% $$
%

\section{Conclusions}
\label{S4}

In the present work, we employed the angular streaking of XUV
ionization of the \H molecule to determine the streaking phase and
time delay corresponding to various orientations of the inter-nuclear
axis relative to the polarization axis of ionizing radiation. The ASX
technique was originally developed to characterize isolated attosecond
pulses from XFEL source on the shot-to-shot basis. This technique was
adapted to determine the streaking phase and applied in our previous
work \cite{Kheifets2022} to the atomic hydrogen. In the present work
we expand this technique to diatomic homonuclear molecules. We
converted the streaking phase to the atomic time delay and found it in
good agreement with our earlier RABBITT simulations
\cite{Serov2017}. Unlike RABBITT, which requires an accurate and
stable synchronization of the ionizing XUV and probing IR pulses,
ASX can determine the streaking phase and time delay from a single XUV
shot. This is essential in XFEL sources with their inherent time
jitter. 

As in earlier works
\cite{PhysRevA.90.013423,PhysRevA.93.063417,Serov2017} we observe a
strong orientation dependence of the molecular time delay.  In most
cases, $\t_a$ remains negative in H, \H and \Hp due to a large
negative CC component. However, $\t_a$ becomes positive in \H and \Hp
in the parallel orientation $\R\parallel\bm e$. This happens when the
photoelectron in the dominant $p$-wave becomes trapped in the
molecular potential well. From the condition of this trapping we can
estimate the depth of this well $U_{\rm eff}$.

While the streaking phase retrieval by ASX was demonstrated for a
diatomic homo-nuclear molecule \H, the proposed method should work for
arbitrary molecular targets. Its application in XFEL will be
particularly beneficial for studying inner shell ionization in atomic
and molecular targets which cannot be ionized at present with
conventional  laser HHG sources.
 
\paragraph*{Acknowledgment:} We thank Rickson Wielian for reviewing the literature
 and useful discussions.  This work is supported by the Discovery
 grant DP190101145 of the Australian Research Council.  Resources of
 National Computational Infrastructure facility (NCI Australia) have
 been employed.

\np
%\bibliography{references,areferences,dreferences,hreferences,rreferences,wreferences,2references,mypapers,reft,ref_1,ureferences,areferences,hreferences,Hreferences,citations,xreferences,greferences}

\end{document}